\documentclass[11pt]{article}
\usepackage[a4paper,margin=2.25cm,headheight=14pt]{geometry}
\usepackage[T1]{fontenc}
\usepackage[utf8]{inputenc}
\usepackage{mathptmx}
\usepackage[english]{babel}
\usepackage{microtype}
\usepackage{setspace}
\usepackage{booktabs}
\usepackage{longtable}
\usepackage{tabularx}
\usepackage{array}
\usepackage{threeparttable}
\usepackage{makecell}
\usepackage{multirow}
\usepackage{pdflscape}
\usepackage{graphicx}
\usepackage{float}
\usepackage[font=small,labelfont=bf]{caption}
\usepackage{subcaption}
\usepackage{siunitx}
\usepackage{xcolor}
\definecolor{navy}{HTML}{17324D}
\definecolor{bluegray}{HTML}{496A81}
\definecolor{teal}{HTML}{2C7A7B}
\definecolor{gold}{HTML}{B8892D}
\definecolor{rose}{HTML}{A55252}
\definecolor{pale}{HTML}{F3F5F7}
\usepackage{tikz}
\usetikzlibrary{arrows.meta,positioning,shapes.geometric,fit,backgrounds,calc}
\usepackage{amsmath,amssymb,mathtools}
\usepackage{enumitem}
\setlist{nosep,leftmargin=*}
\usepackage{csquotes}
\usepackage[backend=biber,style=authoryear,sorting=nyt,maxcitenames=2,maxbibnames=99,doi=true,url=true,isbn=false]{biblatex}
\usepackage{xurl}
\usepackage{hyperref}
\hypersetup{colorlinks=true,linkcolor=navy,citecolor=teal,urlcolor=bluegray,
  pdftitle={LPG Subsidy Reform in Bolivia: Distributional Microsimulation and an ML-ABM Architecture},
  pdfauthor={Ricardo Alonzo Fernández Salguero},
  pdfsubject={Microsimulation, machine learning, and agent-based policy modeling},
  pdfkeywords={Bolivia, LPG, energy subsidies, microsimulation, machine learning, agent-based model, poverty}}
\usepackage{fancyhdr}
\usepackage{titlesec}
\titleformat{\section}{\Large\bfseries\color{navy}}{\thesection.}{0.6em}{}
\titlespacing*{\section}{0pt}{1.5em}{0.8em}
\newcolumntype{Y}{>{\raggedright\arraybackslash}X}
\newcolumntype{C}{>{\centering\arraybackslash}X}
\newcommand{\source}[1]{\par\vspace{1mm}{\footnotesize\textit{Source:} #1}}
\newcommand{\note}[1]{\par\vspace{1mm}{\footnotesize\textit{Note:} #1}}

\begin{document}

\begin{titlepage}
\centering
\vspace*{1.0cm}
{\color{navy}\rule{\textwidth}{1.2pt}}\\[0.8cm]
{\Huge\bfseries LPG Subsidy Reform in Bolivia:\par}
\vspace{0.25cm}
{\LARGE\bfseries distributional microsimulation and an ML--ABM architecture for targeted compensation design\par}
\vspace{0.8cm}
{\Large Ricardo Alonzo Fernández Salguero\par}
\vspace{0.2cm}
{\large Universitat Politècnica de Catalunya -- BarcelonaTech (UPC), Barcelona, Spain\par}
\vspace{0.15cm}
{\large \href{mailto:ricardo.alonzo.fernandez@upc.edu}{ricardo.alonzo.fernandez@upc.edu}\par}
\vspace{0.15cm}
{\large ORCID: \href{https://orcid.org/0000-0002-4189-961X}{0000-0002-4189-961X}\par}
\vspace{0.15cm}
{\large DOI: \href{https://doi.org/10.5281/zenodo.21287432}{10.5281/zenodo.21287432}\par}
\vfill
\begin{minipage}{0.92\textwidth}
\small
\textbf{Abstract.} This study evaluates reform of Bolivia's household liquefied petroleum gas (LPG) subsidy through an integrated architecture of distributional microsimulation, machine learning, and agent-based modeling. The empirical system harmonizes the 2024 Household Survey, the 2015--2016 Family Budget Survey, the 2023 Demographic and Health Survey, monthly hydrocarbon and natural-gas series, and selected macroeconomic indicators. Machine-learning models estimate household LPG demand, food-insecurity vulnerability, current pregnancy, cooking-fuel choice, and latent household archetypes. These learned propensities enter a monthly behavioral model that represents price shocks, liquidity constraints, administrative exclusion, supply scarcity, inertia, hysteresis, and pressure to switch toward solid fuels. The central results show that uncompensated subsidy removal maximizes fiscal savings but increases poverty, extreme poverty, food insecurity, and modeled switching pressure. Full-gap cash compensation for households in income quintiles 1 and 2 preserves substantial fiscal savings and dominates an equivalent voucher when LPG is physically available. An integrated design that adds mother-child transfers and clean-energy transition kits further reduces social damage, although at a higher fiscal cost. Accounting and poverty outcomes are direct ex ante simulations; fuel transitions and health exposures are calibrated projections rather than identified causal effects. Placebo tests, randomization inference, adversarial scenarios, household-level optimization, and global sensitivity analysis define the model's robustness and uncertainty.

\vspace{0.4cm}

\vspace{0.4cm}
\textbf{How to cite:} Fernández Salguero, R. A. (2026). \textit{LPG Subsidy Reform in Bolivia: Distributional Microsimulation and an ML--ABM Architecture for Targeted Compensation Design}. DOI: \href{https://doi.org/10.5281/zenodo.21287432}{10.5281/zenodo.21287432}.
\end{minipage}
\vfill
{\color{navy}\rule{\textwidth}{1.2pt}}
\end{titlepage}

\section{Policy problem, contribution, and scope of the evidence}

Bolivia's generalized LPG subsidy combines several objectives that should be separated analytically: containing the cost of cooking, protecting households with limited liquidity, sustaining an administered consumer price, and discouraging social unrest in a market exposed to supply constraints and diversion. The official price of a 10-kilogram household cylinder is BOB~22.50, which is the baseline used throughout the simulation \parencite{YPFB2026}. A reform can generate large fiscal savings, but fiscal savings are not equivalent to social welfare. A higher cylinder price reduces real disposable income, may crowd out food and health expenditure, and may induce households to combine fuels or return partially to firewood, dung, or other solid fuels. The energy-ladder literature shows that fuel choice is rarely a one-way technological transition: households stack fuels according to prices, availability, risk, taste, and liquidity \parencite{Masera2000}. Household combustion of polluting fuels is associated with health risks, especially for women and children, although the present model estimates exposure pressure rather than clinically caused cases \parencite{WHO2014}.

The institutional question is not only how much to compensate, but also through which instrument. Experimental evidence shows that cash, food, and vouchers can differ in cost, usage, and welfare, and that their relative performance depends on implementation conditions and the desired balance between flexibility and paternalism \parencite{Hidrobo2014,Cunha2014}. International experience with energy-subsidy reform also shows that transfers are effective only when registries, payment systems, sequencing, communication, and administrative coverage work before the price adjustment is imposed \parencite{Mukherjee2023}. A cylinder voucher is therefore not automatically superior to cash. It may be justified when a physically scarce quantity must be rationed, redemption must be traced, or resale must be controlled. When LPG is available and the binding constraint is household purchasing power, a transfer delivered through an existing channel is normally less costly and less administratively fragile.

This study provides an ex ante evaluation that links household microdata to monthly supply information inside an ML--ABM architecture. The objective is not to produce a single deterministic forecast, but to identify which conclusions are accounting identities, which depend on predictive learning, and which are conditional on behavioral calibration. The principal agent is the household observed in the 2024 Household Survey and expanded with survey weights. LPG demand is learned from the Family Budget Survey; current-pregnancy probabilities and child-risk cells are informed by the Demographic and Health Survey; supply scarcity is linked to monthly production, refinery, sales, and network-gas series; and macroeconomic context is retained through World Development Indicators \parencite{INE2024EH,INE2019EPF,INE2025EDSA,INE2026Hydro,WorldBankWDI2026}.

The analysis distinguishes four levels of inference. Fiscal cost identities, disposable-income losses, and poverty reclassification are mechanical results conditional on the assumed price and benefit rules. Demand, vulnerability, and fuel-choice propensities are out-of-sample predictions. Responses to prices outside the observed household range, hysteresis, and solid-fuel switching are calibrated mechanisms subjected to sensitivity analysis. Child-health outcomes are population-risk bridges rather than causal treatment effects. This hierarchy is essential because the flexibility of an agent-based model can otherwise create false confidence.

\begin{table}[H]
\centering
\caption{Evidence hierarchy and defensible strength of claims}
\label{tab:evidence}
\begin{threeparttable}
\begin{tabularx}{\textwidth}{p{3.0cm}Yp{3.0cm}p{3.5cm}}
\toprule
Component & Operation & Defensible strength & Principal restriction \\
\midrule
Gross fiscal savings & Price differential multiplied by imputed annual quantity & High, conditional on price and quantity & The market-equivalent price is a scenario, not a forecast \\
Poverty and extreme poverty & Per-capita income recomputed after the shock and compensation & High as static microsimulation & No general equilibrium or labor-market response \\
Cylinder demand & Family Budget Survey prediction transferred to Household Survey agents & Moderate in aggregates & Limited individual predictive power \\
Food insecurity & Calibrated ensemble with department holdout & Moderate & Cross-sectional data and self-reported outcomes \\
Fuel switching & Learned affinity plus calibrated elasticity, scarcity, and hysteresis & Conditional & Insufficient exogenous price variation \\
Child exposure & Demographic-health risk-cell matching & Exploratory & Not equivalent to causal clinical incidence \\
Policy optimization & Household-instrument selection under a budget & Conditional on welfare weights & Normative weights must remain explicit \\
\bottomrule
\end{tabularx}
\begin{tablenotes}\footnotesize
\item A simulated value can be internally exact while remaining externally unidentified. The hierarchy governs all interpretations below.
\end{tablenotes}
\end{threeparttable}
\end{table}

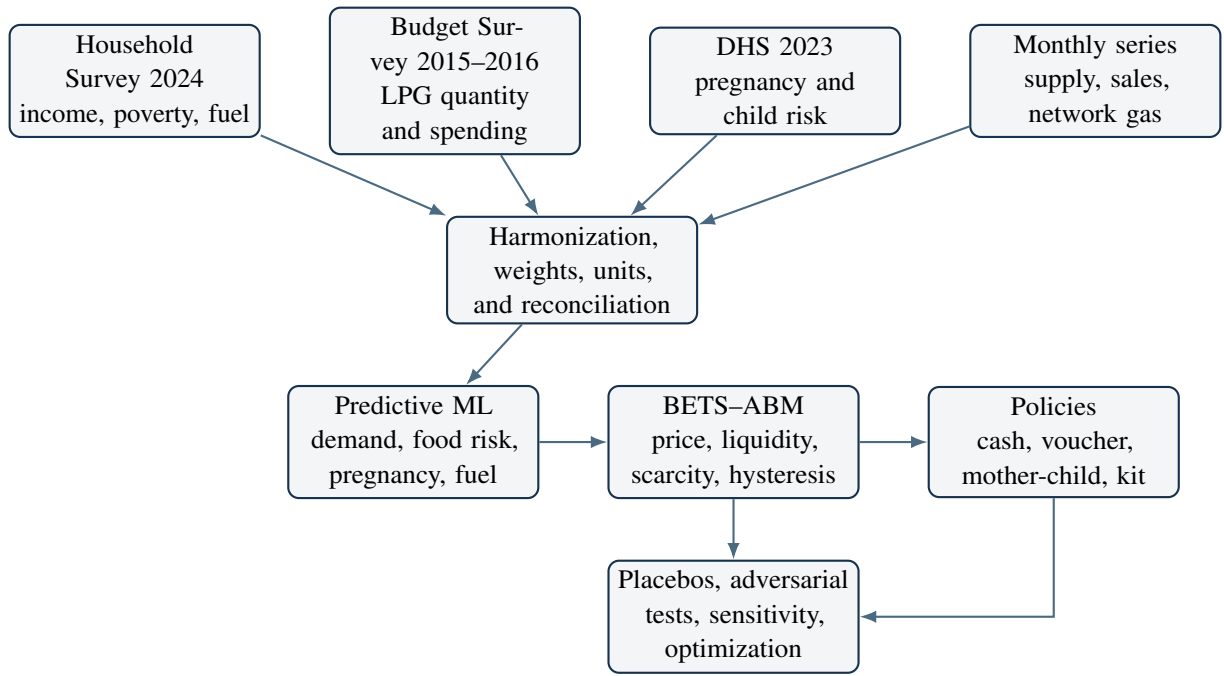
\begin{figure}[H]
\centering
\begin{tikzpicture}[node distance=8mm and 9mm, every node/.style={font=\small}, box/.style={rounded corners,draw=navy,thick,fill=pale,align=center,minimum height=11mm,text width=3.05cm}, arr/.style={-{Latex[length=2.2mm]},thick,draw=bluegray}]
\node[box] (eh) {Household Survey 2024\\income, poverty, fuel};
\node[box,right=of eh] (epf) {Budget Survey 2015--2016\\LPG quantity and spending};
\node[box,right=of epf] (edsa) {DHS 2023\\pregnancy and child risk};
\node[box,right=of edsa] (hydro) {Monthly series\\supply, sales, network gas};
\node[box,below=of epf,xshift=1.55cm] (harm) {Harmonization, weights, units, and reconciliation};
\node[box,below=of harm,xshift=-2.1cm] (ml) {Predictive ML\\demand, food risk, pregnancy, fuel};
\node[box,right=of ml] (abm) {BETS--ABM\\price, liquidity, scarcity, hysteresis};
\node[box,right=of abm] (pol) {Policies\\cash, voucher, mother-child, kit};
\node[box,below=of abm] (rob) {Placebos, adversarial tests, sensitivity, optimization};
\draw[arr] (eh) -- (harm); \draw[arr] (epf) -- (harm); \draw[arr] (edsa) -- (harm); \draw[arr] (hydro) -- (harm);
\draw[arr] (harm) -- (ml); \draw[arr] (ml) -- (abm); \draw[arr] (abm) -- (pol); \draw[arr] (pol) |- (rob); \draw[arr] (abm) -- (rob);
\end{tikzpicture}
\caption{Integrated analytical architecture}
\label{fig:architecture}
\source{Author's design. ABM documentation follows the structural-transparency logic of ODD and ODD+D \parencite{Grimm2020,Muller2013}.}
\end{figure}

\section{Data harmonization and ML--ABM methodology}

The 2024 Household Survey contributes 12718 household observations representing approximately 3.95 million households, with variables on income, poverty, composition, cooking fuel, housing conditions, digital access, and food insecurity \parencite{INE2024EH}. The 2015--2016 Family Budget Survey identifies 5939 LPG-using households representing approximately 1.54 million households and permits construction of annual cylinder demand from the specific product code and reported expenditure \parencite{INE2019EPF}. The 2023 Demographic and Health Survey supplies women- and child-level information used to estimate current pregnancy and health-development risk cells \parencite{INE2025EDSA}. Monthly hydrocarbon tables provide gross gas and liquid production, refinery products, domestic LPG sales, natural gas distributed through networks, and corresponding indices \parencite{INE2026Hydro}. World Development Indicators enter only as macroeconomic context rather than microeconomic identification \parencite{WorldBankWDI2026}.

\begin{table}[H]
\centering
\caption{Harmonized data inventory}
\label{tab:data}
\begin{tabularx}{\textwidth}{p{3.1cm}r r p{2.5cm}Y}
\toprule
Source & Records & Weighted units & Analytical unit & Function \\
\midrule
Household Survey 2024 & 12718 & 3947690 & Households & Poverty, fuel, food insecurity, and household agents \\
Family Budget Survey 2015--2016 & 5939 & 1535690 & LPG households & Observed quantity and expenditure; demand training \\
DHS 2023, children & 88 & 5529 & Risk cells & Child-exposure bridge \\
DHS 2023, women & 13097 & -- & Women & Current-pregnancy probability \\
Hydrocarbon series & 216 & -- & Months & Supply, domestic sales, and scarcity state \\
WDI Bolivia & 66 & -- & Years & Macroeconomic context \\
\bottomrule
\end{tabularx}
\source{Author's harmonization using official microdata and statistical tables \parencite{INE2024EH,INE2019EPF,INE2025EDSA,INE2026Hydro,WorldBankWDI2026}.}
\end{table}

The pipeline preserves the survey weight of each source and avoids pretending that households from different surveys form a longitudinal panel. Instead, it transfers models across sources. LPG demand is learned in the Family Budget Survey and predicted for Household Survey agents. Pregnancy probabilities are learned in the Demographic and Health Survey and aggregated to compatible household age bands. Child risk is assigned through cells defined by wealth, area, age, and fuel. Monthly scarcity enters as an aggregate environmental state. This is more transparent than constructing an individual pseudo-panel with no observational basis.

Annual household LPG demand is represented as
\begin{equation}
q_i=f_{\theta}(X_i)+\varepsilon_i,
\end{equation}
where $X_i$ includes household size, children, women of reproductive age, income, area, department, and housing characteristics. Ridge, Elastic Net, Random Forest, Extra Trees, and histogram gradient boosting are compared. Random Forest and Extra Trees exploit randomized tree ensembles \parencite{Breiman2001}, while histogram gradient boosting belongs to the functional boosting family \parencite{Friedman2001}. Validation holds departments out of sample, limiting geographic leakage. The selected ensemble is rescaled to reproduce the weighted mean observed in the Family Budget Survey.

\begin{table}[H]
\centering
\caption{Out-of-sample performance for annual household LPG demand}
\label{tab:demandmodels}
\begin{tabular}{l S[table-format=1.3] S[table-format=1.3] S[table-format=1.3] l}
\toprule
Model & {RMSE} & {MAE} & {$R^2$} & Validation \\
\midrule
Elastic Net & 6.078 & 4.024 & 0.074 & GroupKFold by department \\
Ridge & 6.078 & 4.105 & 0.074 & GroupKFold by department \\
Histogram gradient boosting & 6.097 & 4.088 & 0.068 & GroupKFold by department \\
Random Forest & 6.106 & 4.078 & 0.065 & GroupKFold by department \\
Extra Trees & 6.137 & 4.040 & 0.056 & GroupKFold by department \\
Weighted mean benchmark & 6.318 & 4.909 & -0.001 & Constant weighted benchmark \\
\bottomrule
\end{tabular}
\note{The predictive gain over the benchmark is genuine but modest; demand estimates are more defensible in aggregates than as exact household forecasts.}
\end{table}

\begin{figure}[H]
\centering
\includegraphics[width=0.78\textwidth]{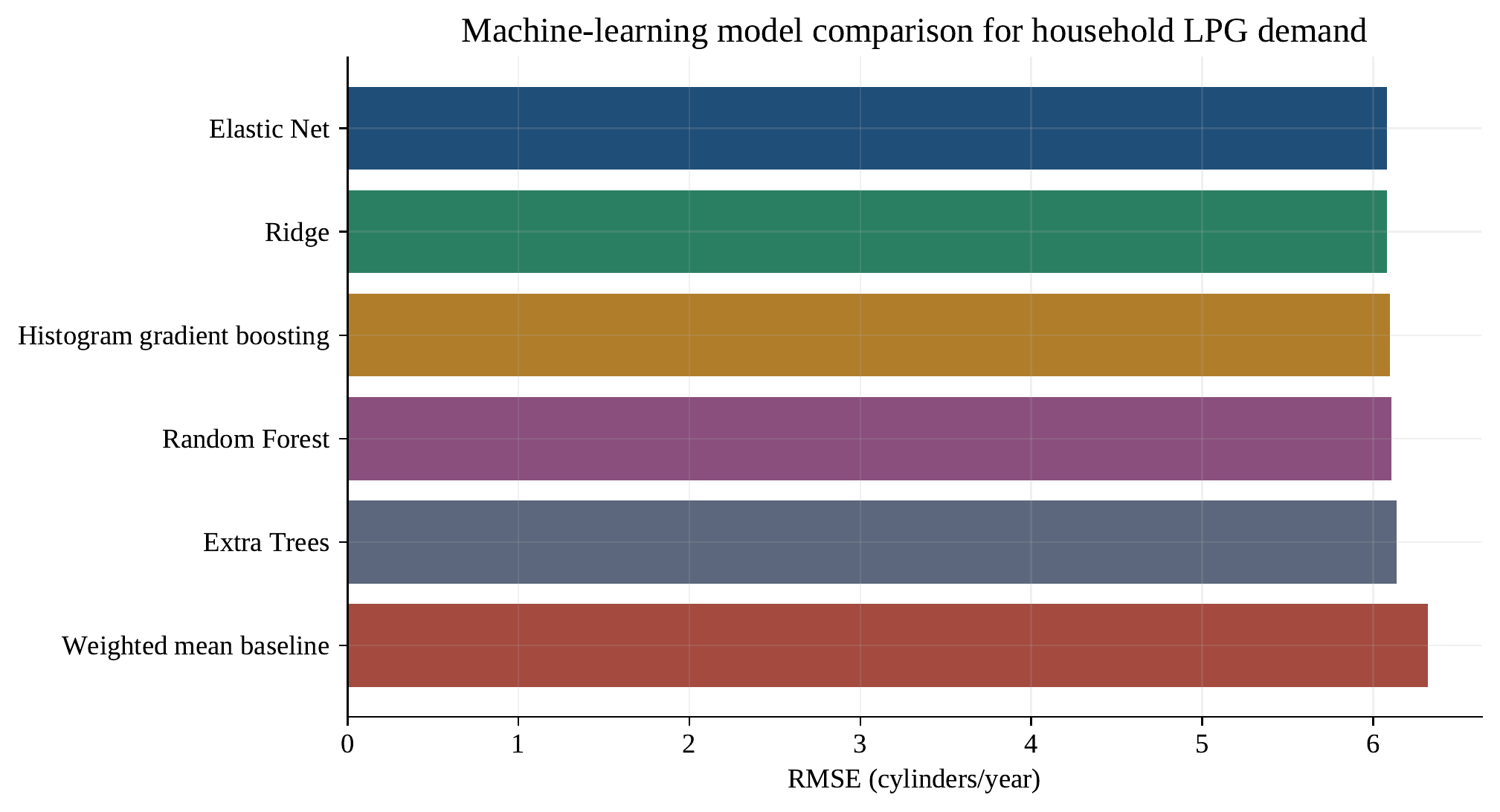}
\caption{Machine-learning comparison for household LPG demand}
\label{fig:demand}
\source{Author's calculations using the 2015--2016 Family Budget Survey.}
\end{figure}

The demand model reproduces the weighted mean of 15.12 cylinders per year, but compresses the observed distribution. The imputed median is 14.76 compared with 12 observed cylinders, while the imputed 90th percentile is 17.32 compared with 24 observed cylinders. This regression toward the center understates intensive users and is therefore carried forward as a limitation and a sensitivity dimension rather than hidden by the ensemble.

\begin{table}[H]
\centering
\caption{Reconciliation of observed and imputed LPG demand}
\label{tab:reconciliation}
\begin{tabular}{l S[table-format=2.3] S[table-format=2.3]}
\toprule
Statistic & {Observed in budget survey} & {Imputed in household agents} \\
\midrule
Weighted mean & 15.123 & 15.123 \\
Weighted median & 12.000 & 14.756 \\
10th percentile & 12.000 & 13.303 \\
90th percentile & 24.000 & 17.318 \\
Share with 12 or more & 0.9998 & 1.0000 \\
\bottomrule
\end{tabular}
\source{Author's calculations using the Family Budget Survey and Household Survey.}
\end{table}

Behavioral probabilities are trained without using an outcome as its own predictor. For food insecurity, the three strongest out-of-fold models are averaged and calibrated using weighted isotonic regression. Calibration matters because discrimination does not guarantee reliable probabilities \parencite{Niculescu2005}. Current pregnancy is rare, so accuracy is not an appropriate selection criterion by itself; ROC-AUC, precision-recall AUC, Brier score, and log loss are reported. Cooking-fuel choice is treated as a multicategory problem and compared with multinomial logit.

\begin{table}[H]
\centering
\caption{Selected behavioral models}
\label{tab:behavior}
\begin{tabularx}{\textwidth}{p{3.0cm}p{3.2cm}S[table-format=1.3]S[table-format=1.3]S[table-format=1.3]Y}
\toprule
Outcome & Selected model & {ROC-AUC} & {PR-AUC} & {Brier} & Assessment \\
\midrule
Current pregnancy & Histogram gradient boosting & 0.703 & 0.053 & 0.029 & Moderate discrimination under severe class imbalance \\
Moderate/severe food insecurity & Calibrated top-3 ensemble & 0.674 & 0.368 & 0.176 & Calibrated out-of-sample probabilities \\
Severe food insecurity & Calibrated top-3 ensemble & 0.680 & 0.143 & 0.073 & Lowest Brier score among candidates \\
Cooking-fuel choice & Histogram gradient boosting & \multicolumn{1}{c}{--} & \multicolumn{1}{c}{--} & \multicolumn{1}{c}{0.503} & Accuracy 0.614; macro-F1 0.396 \\
\bottomrule
\end{tabularx}
\end{table}

\begin{figure}[H]
\centering
\begin{subfigure}{0.49\textwidth}
\includegraphics[width=\textwidth]{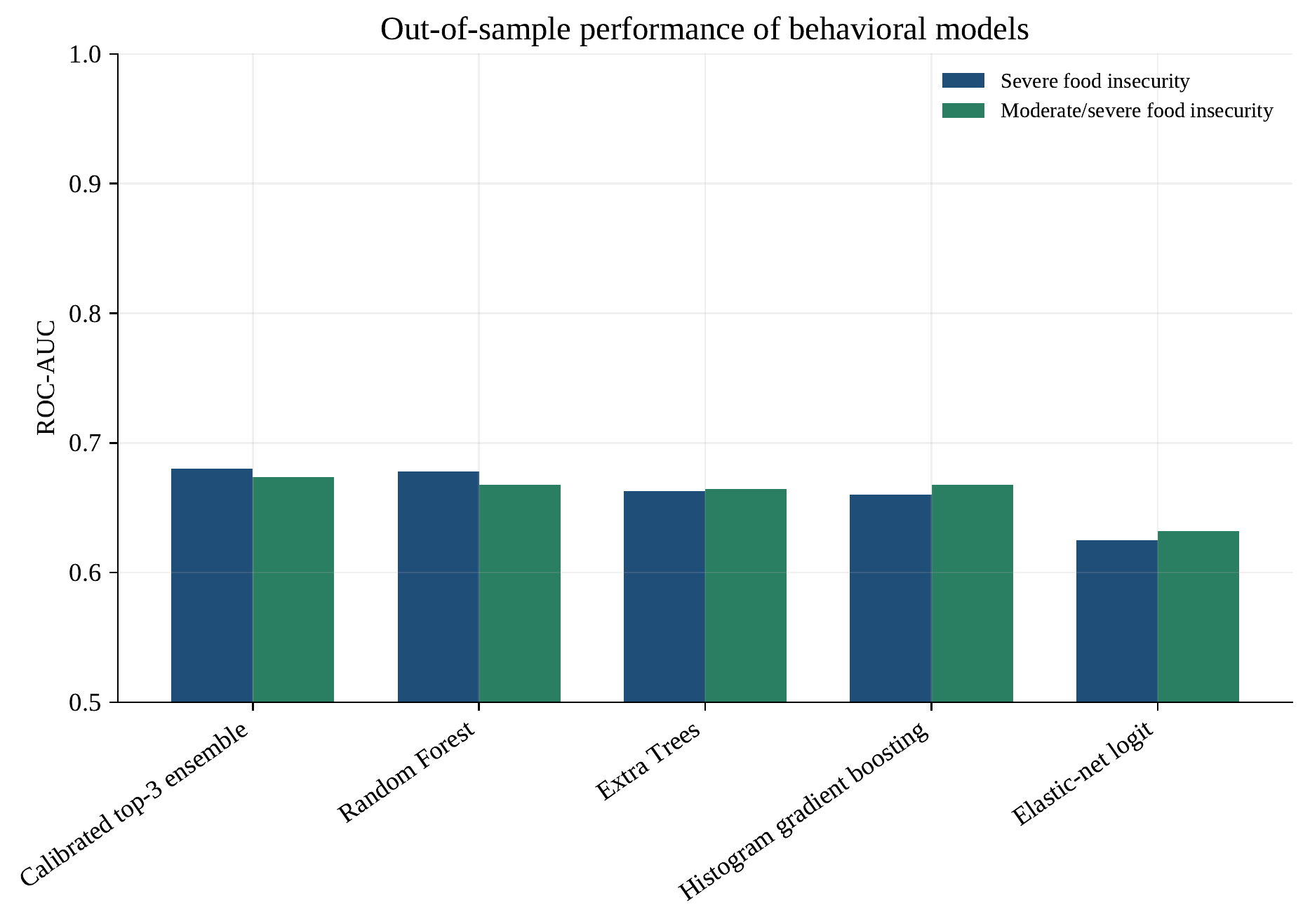}
\caption{Discrimination}
\end{subfigure}\hfill
\begin{subfigure}{0.49\textwidth}
\includegraphics[width=\textwidth]{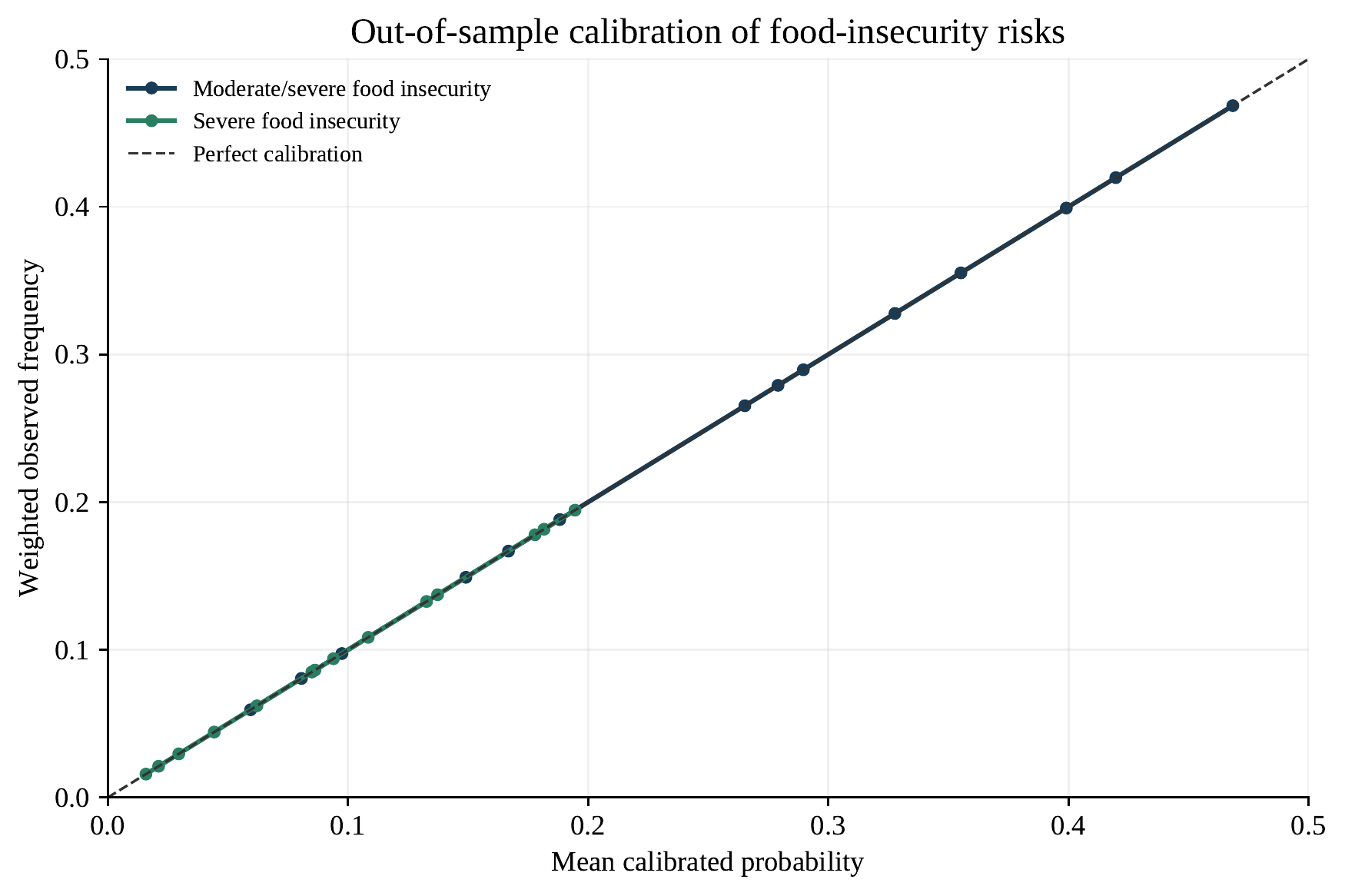}
\caption{Probability calibration}
\end{subfigure}
\caption{Out-of-sample behavioral-model assessment}
\label{fig:behavior}
\source{Author's calculations using the 2024 Household Survey.}
\end{figure}

Learned fuel probabilities are interpreted as affinities rather than price elasticities. K-means clustering over poverty, rurality, fuel, digital access, food risk, and household composition produces household archetypes that enter the ABM as compact heterogeneity states. The archetypes are descriptive: they simplify repeated decision rules while preserving the main gradients across vulnerable rural solid-fuel households, urban network-gas households, LPG-dependent low-income households, and higher-income modern-energy households.

\begin{figure}[H]
\centering
\includegraphics[width=0.82\textwidth]{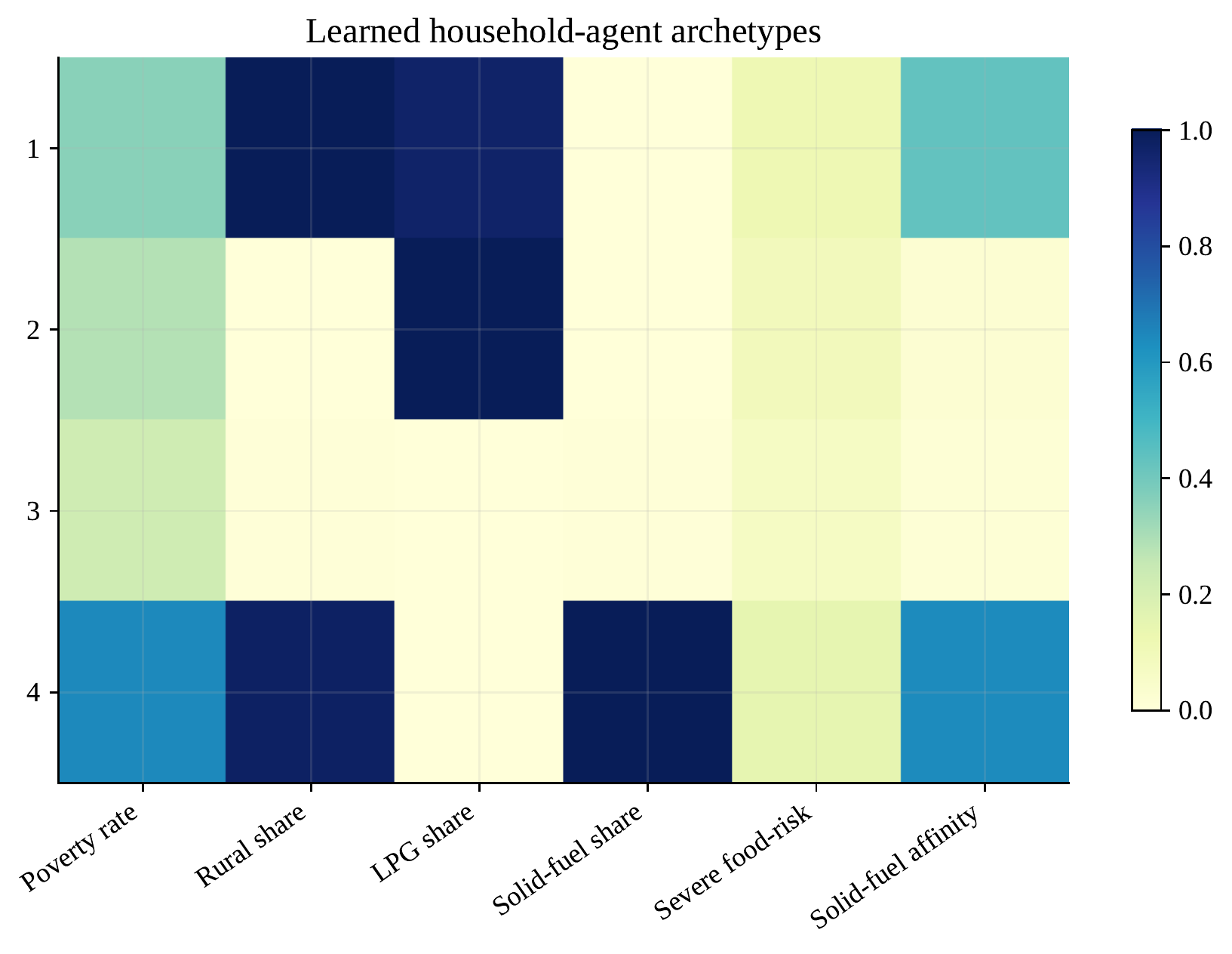}
\caption{Learned household-agent archetypes}
\label{fig:archetypes}
\source{Author's calculations. Cell values are normalized within features for visualization.}
\end{figure}

Monthly LPG supply is evaluated with a temporal holdout. A seasonal-naive benchmark outperforms Ridge, histogram gradient boosting, Extra Trees, and Random Forest. This negative result is retained because model complexity is not evidence of superior forecasting. The archived supply state is therefore anchored to observed monthly information and a transparent seasonal benchmark rather than to an overfit machine-learning forecast.

\begin{table}[H]
\centering
\caption{Temporal holdout for monthly LPG sales}
\label{tab:supplymodels}
\begin{tabular}{l S[table-format=5.0] S[table-format=5.0] S[table-format=1.3] S[table-format=1.2]}
\toprule
Model & {RMSE} & {MAE} & {$R^2$} & {MAPE (\%)} \\
\midrule
Seasonal naive & 17600 & 15206 & 0.317 & 3.16 \\
Ridge & 25431 & 22280 & -0.426 & 4.56 \\
Histogram gradient boosting & 39512 & 33628 & -2.442 & 6.81 \\
Extra Trees & 41175 & 37945 & -2.737 & 7.75 \\
Random Forest & 43234 & 37710 & -3.120 & 7.69 \\
\bottomrule
\end{tabular}
\source{Author's rolling holdout using official monthly hydrocarbon series \parencite{INE2026Hydro}.}
\end{table}

\begin{figure}[H]
\centering
\includegraphics[width=0.86\textwidth]{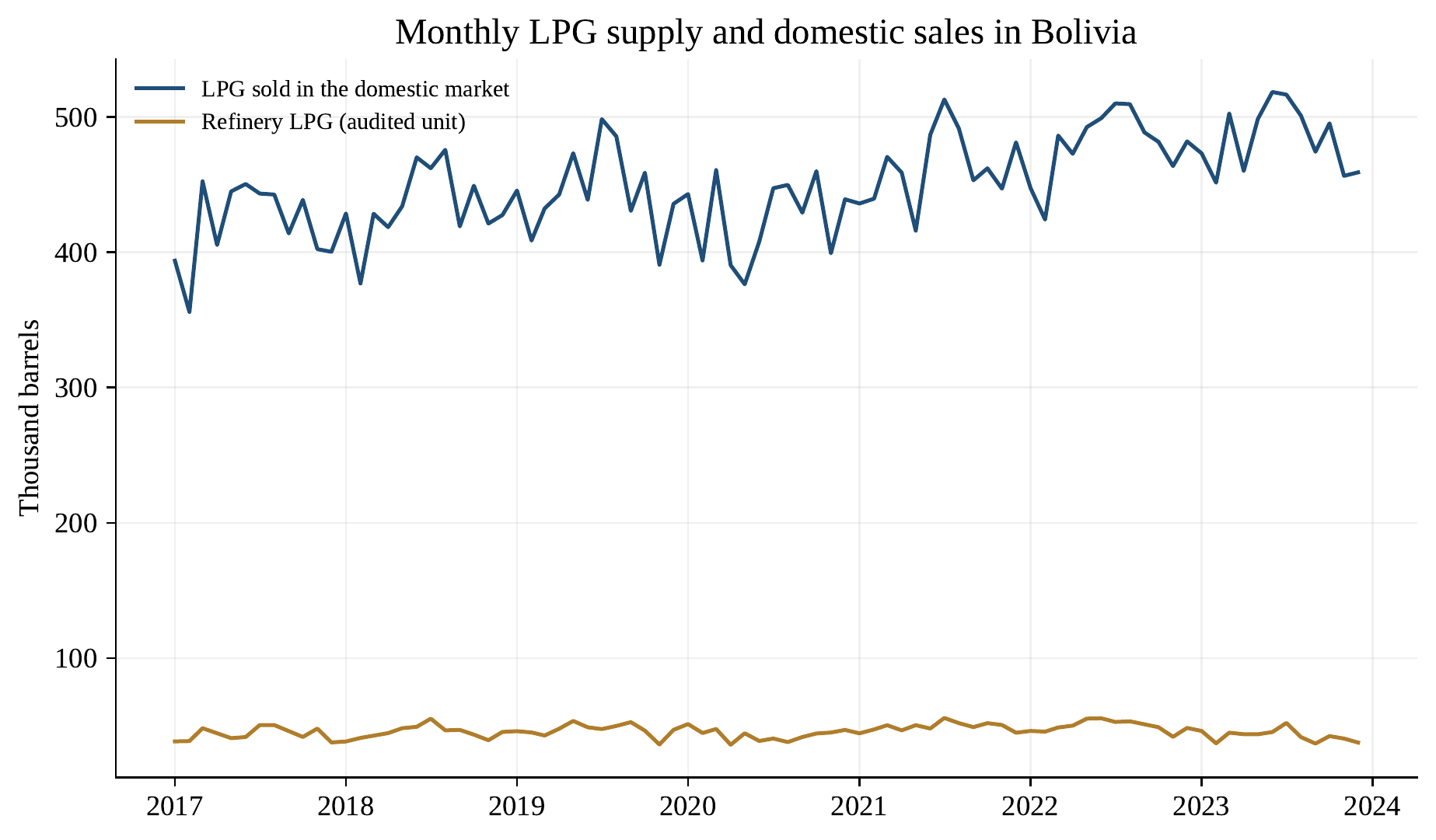}
\caption{Monthly LPG supply and domestic sales in Bolivia}
\label{fig:supply}
\source{Author's processing of official monthly hydrocarbon tables \parencite{INE2026Hydro}.}
\end{figure}

The Behavioral Ensemble Transition Surface, abbreviated BETS, maps learned states and calibrated responses into monthly switching pressure. For household $i$ in month $t$,
\begin{equation}
U_{it}=\frac{\max\{0,\Delta p_t q_i-C_{it}\}}{\max\{\Delta p_t q_i,1\}},
\end{equation}
where $\Delta p_t$ is the price increase, $q_i$ is imputed demand, and $C_{it}$ is effective compensation after take-up and leakage. The latent switching index is
\begin{equation}
L_{it}=\beta_0+\beta_1 U_{it}+\beta_2\frac{\Delta p_t q_i-C_{it}}{Y_i}
+\beta_3 A_i^{S}+\beta_4 R_i^{F}+\beta_5 E_i+\beta_6 S_t+\beta_7 H_{i,t-1},
\end{equation}
where $A_i^{S}$ is learned solid-fuel affinity, $R_i^{F}$ is calibrated food-insecurity risk, $E_i$ is administrative-exclusion risk, $S_t$ is scarcity, and $H_{i,t-1}$ is hysteresis. Monthly pressure is
\begin{equation}
\Pr(Switch_{it}=1)=\operatorname{logit}^{-1}(\lambda L_{it}),
\end{equation}
where $\lambda$ is a scenario elasticity. Learned quantities and calibrated quantities are kept separate in the assumption registry.

Policy accounting distinguishes gross savings, transfers, administration, leakage, remote-delivery costs, and setup investment. For design $d$,
\begin{equation}
NS_d=\sum_i w_i q_i(p_m-p_0)-B_d-A_d-L_d-F_d,
\end{equation}
where $p_0=22.50$, $p_m$ is the simulated market-equivalent price, $B_d$ is direct benefit cost, $A_d$ is administration, $L_d$ is leakage or diversion, and $F_d$ is fixed setup. Household post-policy income is
\begin{equation}
Y_{id}^{*}=Y_i-q_i(p_m-p_0)+T_{id},
\end{equation}
and poverty is recalculated using the survey poverty lines. This accounting makes clear why a voucher and cash transfer with the same nominal entitlement can have different net protection.

\section{Predictive, distributional, fiscal, and dynamic results}

At the official price, the simulated baseline poverty rate is 37.62 percent and extreme poverty is 12.76 percent. At BOB~140 per cylinder, uncompensated reform produces approximately BOB~3.81 billion in net fiscal savings, but raises poverty by 1.37 percentage points and extreme poverty by 1.03 points. A fixed Q1-Q2 transfer retains most savings but absorbs only a small share of the actual price gap. Full-gap Q1-Q2 cash compensation reduces the poverty increase to 0.38 points and the extreme-poverty increase to 0.14 points while preserving about BOB~2.53 billion in net savings. The comparable voucher preserves less fiscal space and leaves greater modeled damage because its take-up, administration, leakage, and setup assumptions are less favorable.

\begin{table}[H]
\centering
\caption{Central policy scenarios at BOB 140 per cylinder}
\label{tab:central}
{\small
\begin{tabularx}{\textwidth}{Yrrrrr}
\toprule
Design & \makecell[r]{Net savings\\(million BOB)} & \makecell[r]{Poverty\\(\%)} & \makecell[r]{Extreme\\poverty (\%)} & \makecell[r]{Solid-fuel pressure\\(thousand)} & \makecell[r]{Incremental\\damage} \\
\midrule
No compensation & 3,807 & 39.00 & 13.79 & 344.5 & 637.3 \\
Fixed cash, Q1--Q2 & 3,616 & 38.94 & 13.60 & 320.8 & 565.0 \\
Full-gap cash, Q1--Q2 & 2,528 & 38.01 & 12.90 & 232.0 & 282.6 \\
Full-gap voucher, Q1--Q2 & 2,474 & 38.07 & 12.96 & 239.3 & 305.6 \\
Cash plus mother-child & 2,333 & 37.93 & 12.76 & 229.7 & 251.5 \\
Integrated design & 2,001 & 37.93 & 12.76 & 198.0 & 200.1 \\
\bottomrule
\end{tabularx}
}
\note{The social-damage index is an explicit weighted combination of poverty, extreme poverty, food insecurity, switching pressure, and child exposure. It is useful for ranking scenarios, not as a natural physical unit.}
\end{table}

\begin{figure}[H]
\centering
\begin{subfigure}{0.49\textwidth}
\includegraphics[width=\textwidth]{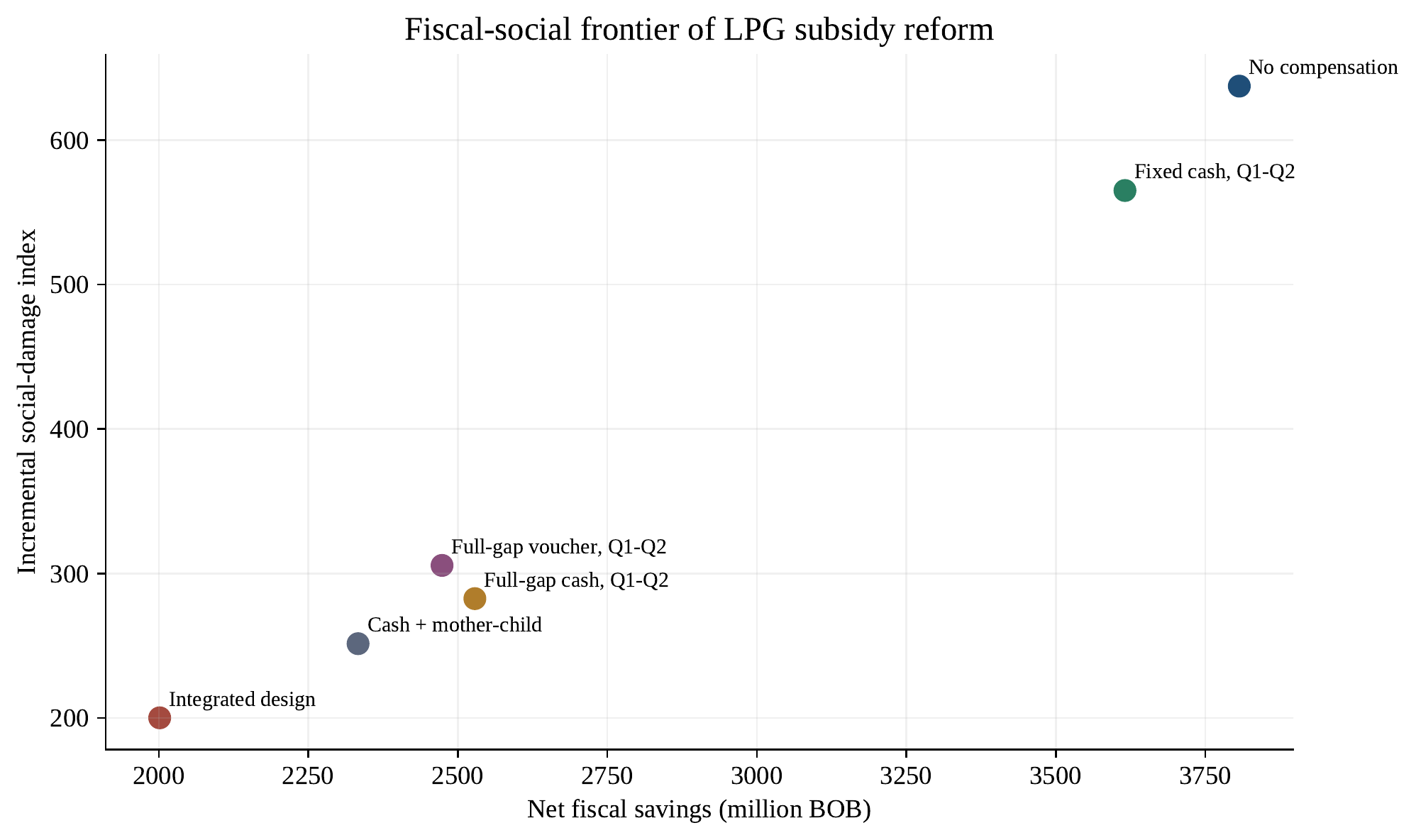}
\caption{Fiscal-social frontier}
\end{subfigure}\hfill
\begin{subfigure}{0.49\textwidth}
\includegraphics[width=\textwidth]{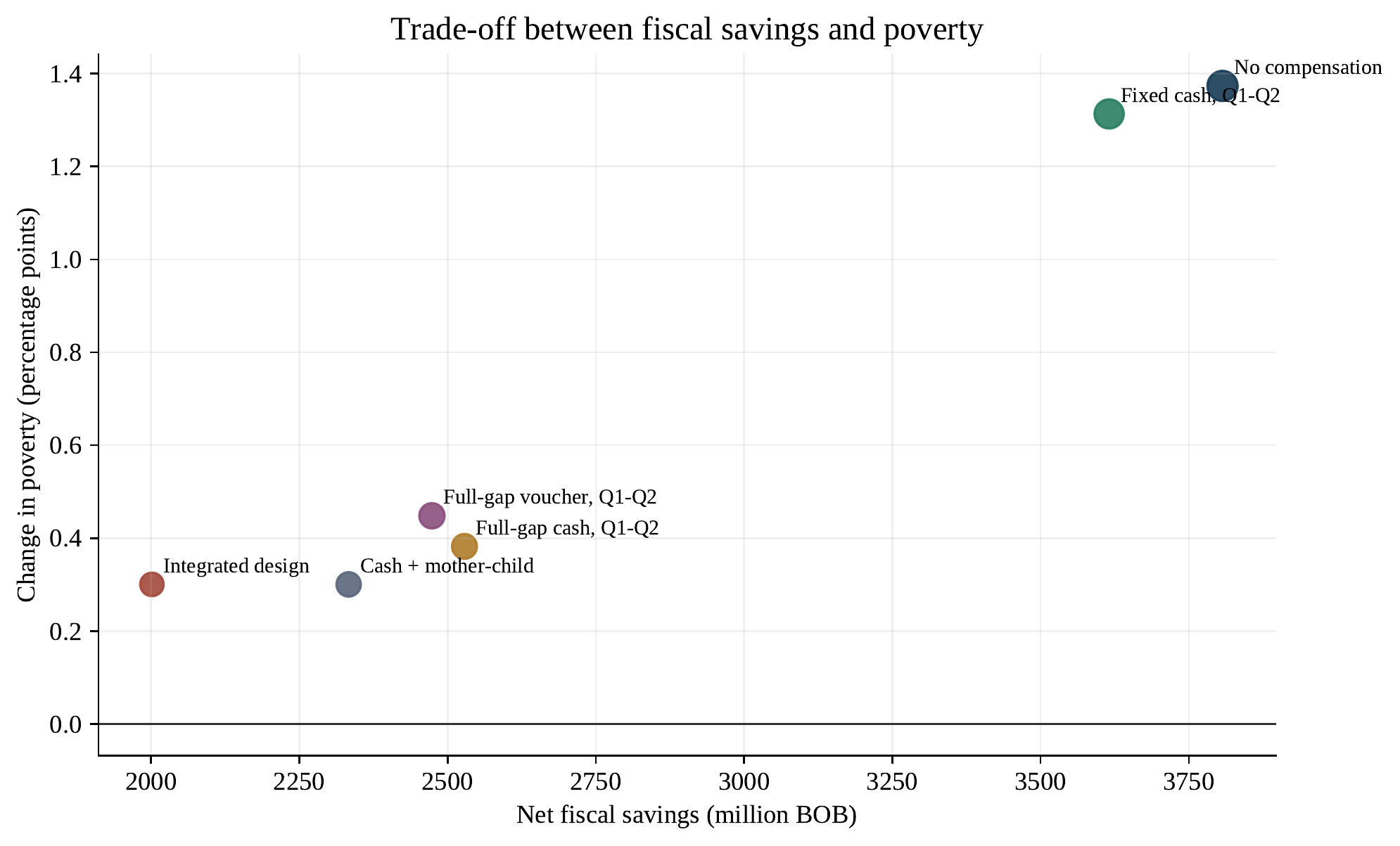}
\caption{Savings-poverty trade-off}
\end{subfigure}
\caption{Central trade-offs across policy designs}
\label{fig:frontier}
\source{Author's microsimulation.}
\end{figure}

The fiscal ranking does not imply that the least costly program is socially efficient. Uncompensated removal has no administrative cost but transfers the full adjustment to households. Full-gap cash has larger direct cost but low setup and administration because it is designed around an existing payment channel. A voucher reduces household exposure but adds redemption infrastructure, verification, clearing, remote-delivery cost, and diversion risk. The integrated design combines full-gap protection, mother-child support, and transition kits for current solid-fuel households; it therefore has the largest protection budget but the lowest modeled damage among the compared designs.

\begin{figure}[H]
\centering
\includegraphics[width=0.88\textwidth]{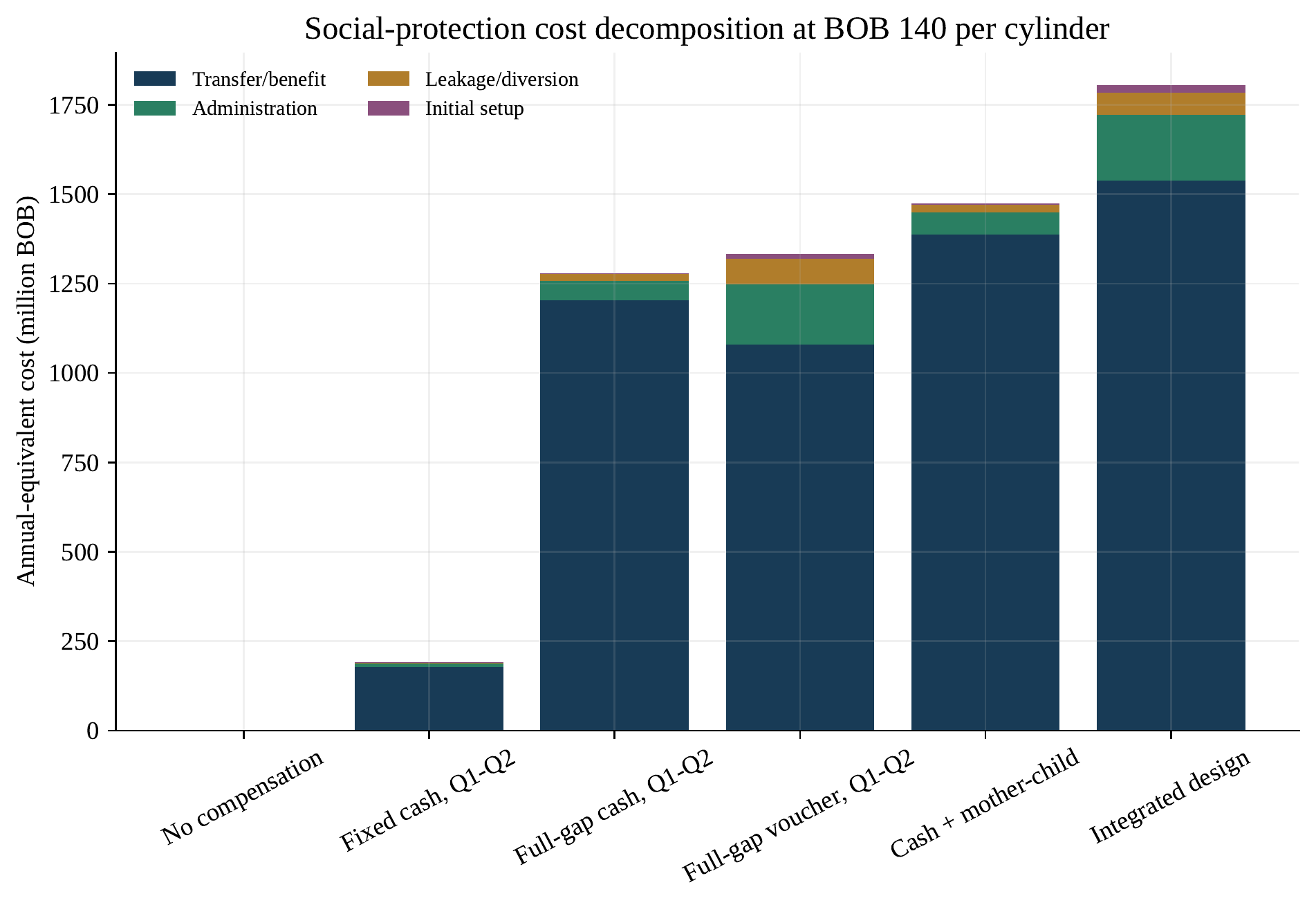}
\caption{Cost decomposition of social-protection alternatives}
\label{fig:costs}
\source{Author's calculations.}
\end{figure}

Distributional incidence is strongly regressive without compensation. At BOB~140, average annual net energy loss is approximately BOB~969 in the first quintile and BOB~1071 in the second quintile. Full-gap cash reduces these values to approximately BOB~136 and BOB~117 because take-up and leakage prevent perfectly complete protection. The voucher leaves larger residual losses in the same quintiles. Higher quintiles remain exposed because the targeted designs deliberately preserve the fiscal adjustment outside Q1-Q2.

\begin{table}[H]
\centering
\caption{Equity and targeting metrics at BOB 140}
\label{tab:equity}
{\small
\begin{tabularx}{\textwidth}{Yrrrrr}
\toprule
Scenario & \makecell[r]{Poor-benefit\\share} & \makecell[r]{Inclusion\\error} & \makecell[r]{Exclusion\\error} & \makecell[r]{Net savings\\(million BOB)} & Damage \\
\midrule
No compensation & 0.000 & 0.000 & 1.000 & 3,807 & 637.3 \\
Full-gap cash, Q1--Q2 & 0.832 & 0.168 & 0.350 & 2,528 & 282.6 \\
Full-gap voucher, Q1--Q2 & 0.832 & 0.168 & 0.350 & 2,474 & 305.6 \\
Integrated design & 0.838 & 0.162 & 0.047 & 2,001 & 200.1 \\
\bottomrule
\end{tabularx}
}
\source{Author's calculations with survey weights.}
\end{table}

\begin{figure}[H]
\centering
\includegraphics[width=0.87\textwidth]{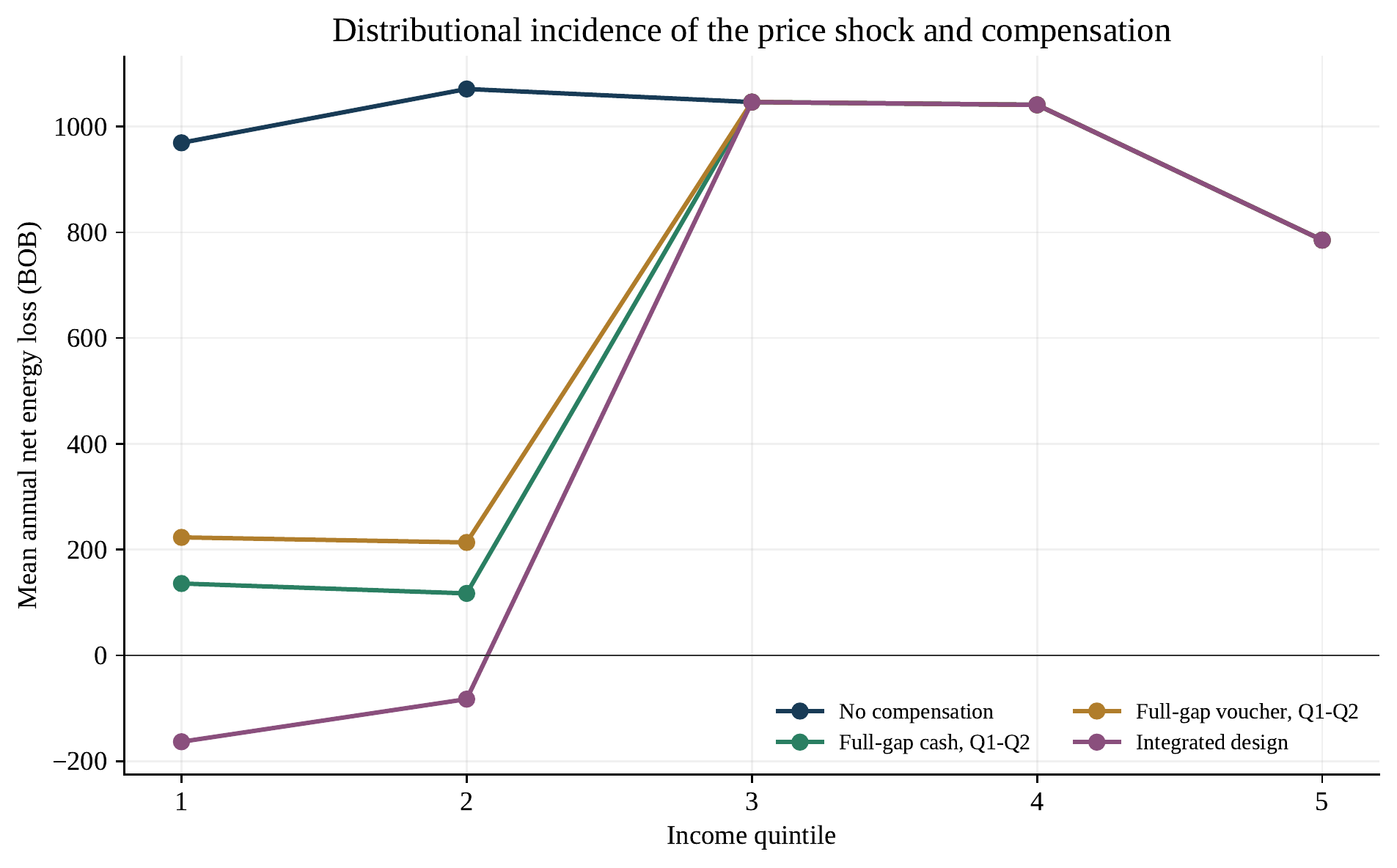}
\caption{Distributional incidence of the price shock and compensation}
\label{fig:incidence}
\source{Author's calculations.}
\end{figure}

The ABM introduces monthly seasonality, scarcity, inertia, and hysteresis. The official-price baseline is not mechanically zero because agents retain a low structural probability of solid-fuel pressure associated with pre-existing vulnerability and seasonal supply variation. Under uncompensated BOB~140 and BOB~200 scenarios, switching pressure rises sharply and remains persistent. Full-gap cash reduces the level but does not eliminate it, because non-Q1-Q2 households, administrative exclusion, and scarcity remain. The integrated design performs best because it protects LPG users and separately addresses households already dependent on solid fuels.

\begin{figure}[H]
\centering
\includegraphics[width=0.89\textwidth]{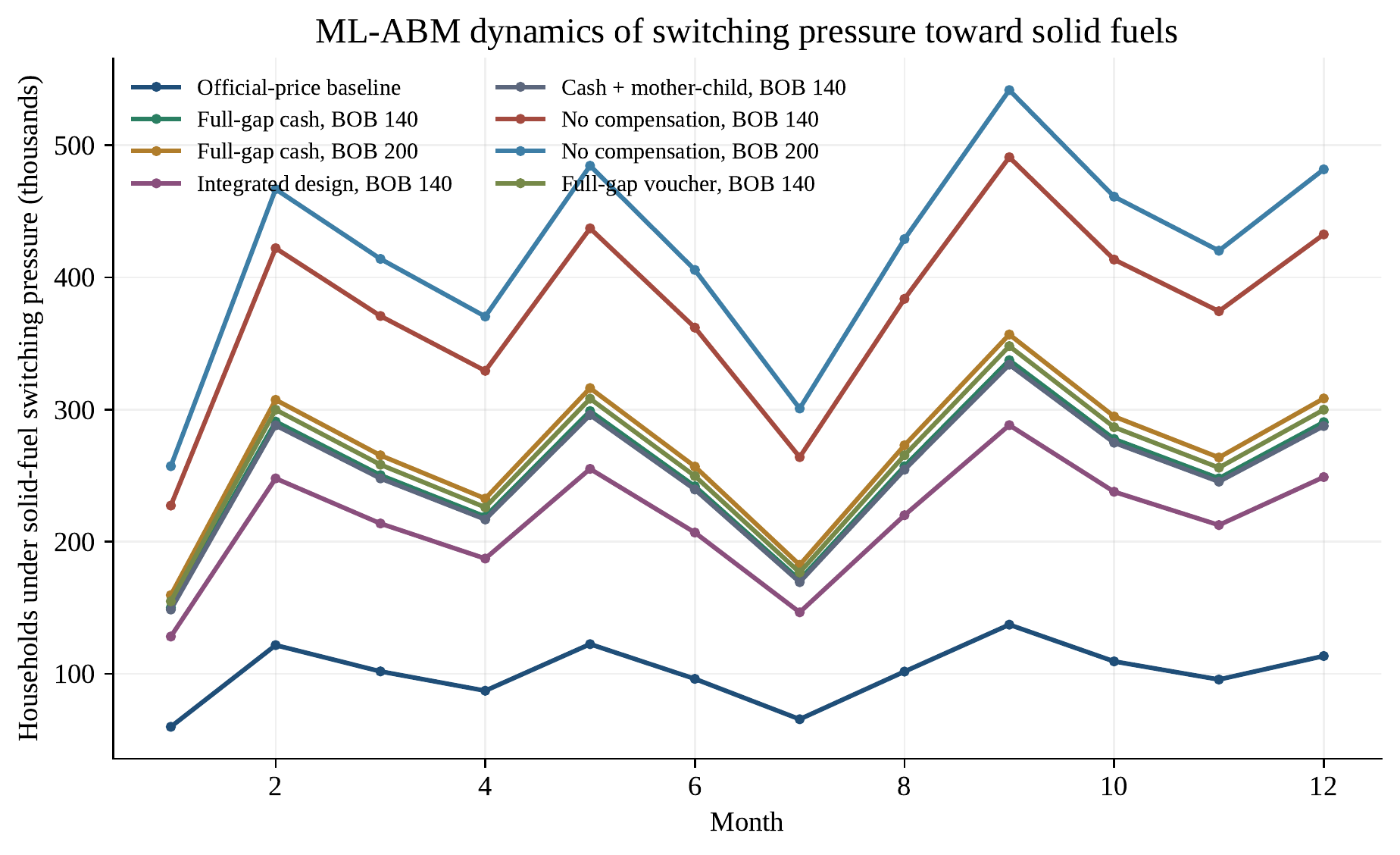}
\caption{Monthly ML--ABM dynamics of pressure to switch toward solid fuels}
\label{fig:dynamics}
\source{Author's BETS--ABM simulation.}
\end{figure}

The household-level optimizer does not choose one universal program. It constructs feasible options for each household, ranks expected social benefit per unit of cost, and recalculates outcomes after every allocation. At BOB~100 million it concentrates on the highest marginal benefit cases. Additional spending expands protection until the eligible portfolio saturates at approximately BOB~690 million. Beyond that point, the same allocation is repeated because no additional option satisfies the eligibility and mutual-exclusivity rules. This saturation is a feature of the constrained portfolio, not evidence that larger budgets are generally useless.

\begin{table}[H]
\centering
\caption{Household-level policy optimizer}
\label{tab:optimizer}
{\small
\begin{tabularx}{\textwidth}{rrrrrr}
\toprule
\makecell{Budget\\(million BOB)} & \makecell{Spent\\(million BOB)} & \makecell{Selected\\options} & \makecell{Net savings\\(million BOB)} & \makecell{Poverty\\(\%)} & Damage \\
\midrule
100 & 100 & 757 & 3,707 & 39.00 & 438.3 \\
250 & 250 & 1,523 & 3,557 & 39.00 & 431.1 \\
500 & 500 & 4,102 & 3,307 & 38.84 & 364.6 \\
750 & 690 & 6,104 & 3,116 & 38.74 & 312.9 \\
1,000 & 690 & 6,104 & 3,116 & 38.74 & 312.9 \\
2,000 & 690 & 6,104 & 3,116 & 38.74 & 312.9 \\
\bottomrule
\end{tabularx}
}
\source{Author's household-option optimization.}
\end{table}

\begin{figure}[H]
\centering
\includegraphics[width=0.82\textwidth]{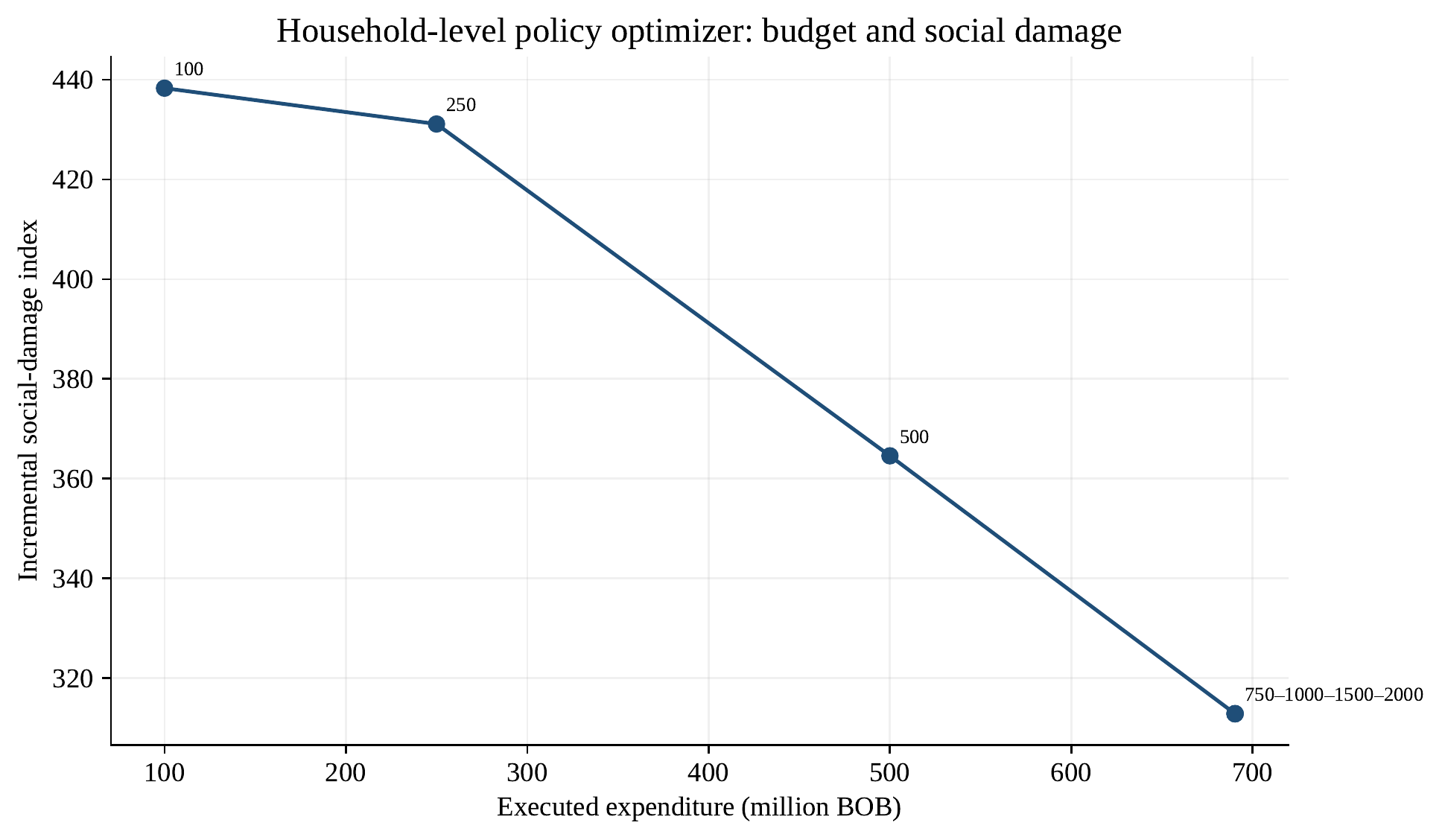}
\caption{Executed policy expenditure and modeled social damage}
\label{fig:optimizer}
\source{Author's calculations. Labels indicate nominal budget ceilings in million BOB.}
\end{figure}

\section{Robustness, falsification, sensitivity, and identification}

ABM credibility requires more than a plausible narrative. The ODD and ODD+D traditions emphasize transparent specification of state variables, scheduling, decision rules, and human choice \parencite{Grimm2020,Muller2013}. Validation should combine empirical pattern reproduction, comparison with simpler models, uncertainty analysis, and falsification rather than rely on face validity alone \parencite{Collins2024}. The present package therefore compares ML with transparent benchmarks, reproduces observed aggregate moments, tests negative exposures, randomizes treatment assignment, evaluates false eligibility, and varies the principal behavioral and administrative assumptions.

The two negative-exposure tests assign the LPG-price mechanism to households connected to network gas or using electricity and other fuels. Their estimated effect is zero because no LPG quantity is imputed to them. The false-eligibility test gives Q4-Q5 households the targeting rule intended for Q1-Q2 and finds no comparable social benefit. Randomization inference repeatedly assigns an equal treated count and compares the observed targeting statistic with its placebo distribution; the resulting $p$-value is 0.0033.

\begin{table}[H]
\centering
\caption{Placebo and refutation tests}
\label{tab:placebos}
\begin{tabularx}{\textwidth}{Y S[table-format=6.1] S[table-format=6.1] S[table-format=1.4] c}
\toprule
Test & {Statistic} & {Reference} & {$p$-value} & Pass \\
\midrule
Negative exposure: network-gas households & 0.0 & 170266.8 & 1.0000 & Yes \\
Negative exposure: electricity/other households & 0.0 & 170266.8 & 1.0000 & Yes \\
Equal-count randomization inference & 170266.8 & 87822.0 & 0.0033 & Yes \\
False eligibility among Q4--Q5 & 0.0 & 96.5 & \multicolumn{1}{c}{--} & Yes \\
\bottomrule
\end{tabularx}
\source{Author's placebo and randomization tests.}
\end{table}

The adversarial grid crosses market prices, take-up, leakage, behavioral elasticity, and scarcity. The main ranking is stable over a broad portion of the grid: full-gap cash normally dominates the equivalent voucher in net savings and modeled damage, while uncompensated removal dominates only on fiscal savings. However, this ranking is conditional on physical availability. If scarcity is severe and the policy objective becomes quantity rationing rather than income replacement, the voucher gains an institutional role that is not captured by a pure welfare ranking.

Global sensitivity uses 600 stratified parameter draws. Rank correlations show that the market-equivalent price almost completely determines gross and net fiscal savings; the generalized income shock dominates poverty; and behavioral elasticity dominates solid-fuel switching and the composite damage index. These results are methodologically important because the largest reported uncertainty comes from parameters not causally identified in the cross-sectional household data. Sensitivity analysis follows the principle that complex models should be evaluated over the joint parameter space rather than through isolated one-at-a-time changes \parencite{Saltelli2010}.

\begin{figure}[H]
\centering
\includegraphics[width=0.84\textwidth]{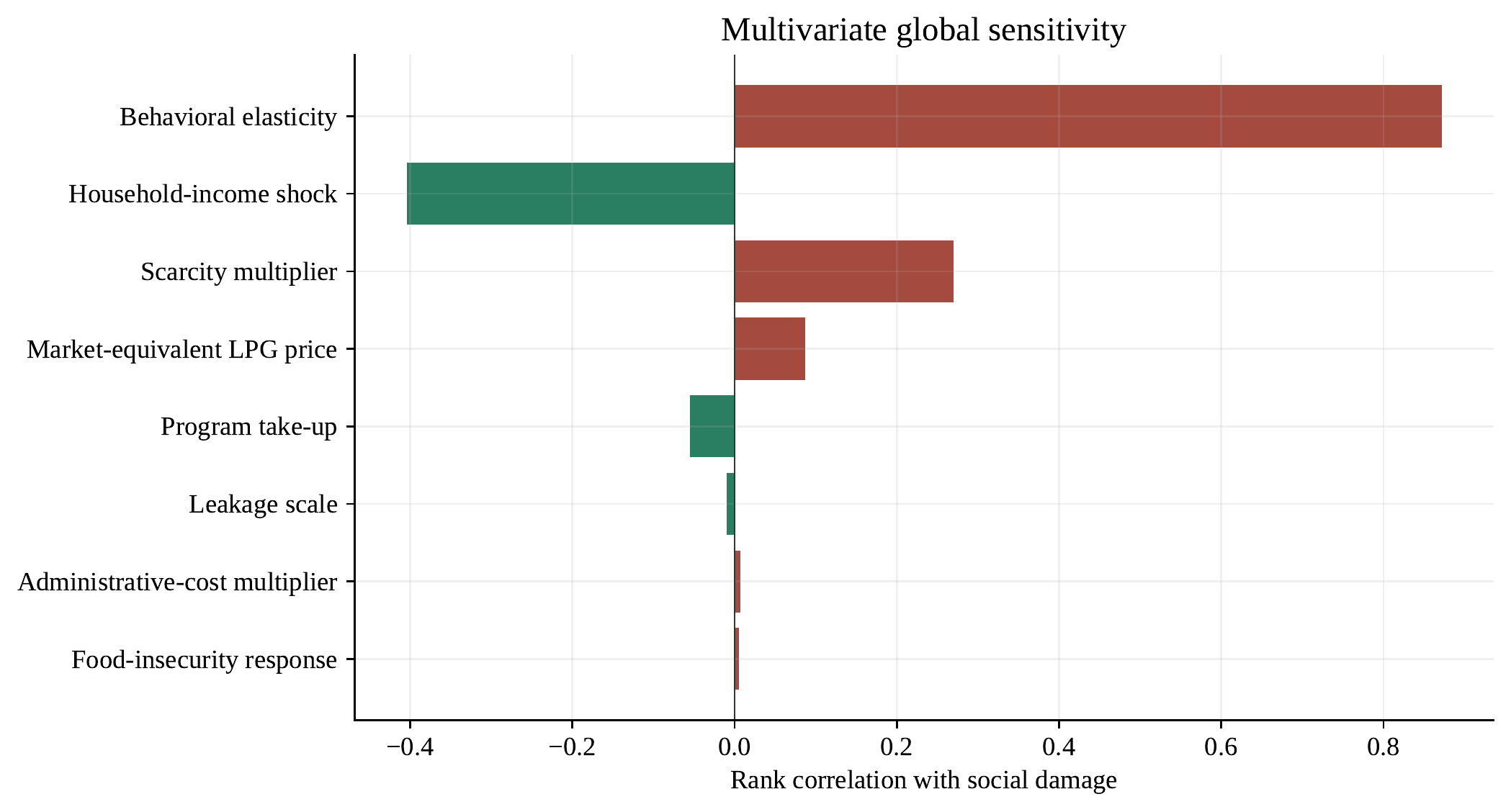}
\caption{Multivariate global sensitivity of the social-damage index}
\label{fig:sensitivity}
\source{Author's sensitivity draws, following global-sensitivity principles \parencite{Saltelli2010}.}
\end{figure}

\begin{table}[H]
\centering
\caption{Dominant parameters by outcome}
\label{tab:sensitivity}
\begin{tabularx}{\textwidth}{p{3.0cm}p{4.0cm}S[table-format=1.3]Y}
\toprule
Outcome & Parameter & {Spearman $\rho$} & Interpretation \\
\midrule
Net fiscal savings & Market-equivalent price & 0.983 & The gap relative to BOB~22.50 determines gross savings \\
Net fiscal savings & Take-up & -0.158 & Greater effective compensation reduces savings \\
Poverty & Generalized income shock & -0.991 & A macroeconomic deterioration amplifies reform effects \\
Social damage & Behavioral elasticity & 0.872 & Solid-fuel transition pressure dominates the index \\
Solid-fuel pressure & Behavioral elasticity & 0.943 & Critical parameter not causally identified \\
\bottomrule
\end{tabularx}
\end{table}

External validity faces five principal limitations. First, the Family Budget Survey and Household Survey refer to different periods; transfer learning preserves observable relationships and selected moments, not immutable preferences. Second, a national household treatment at the simulated price is not observed, so the price elasticity is calibrated. Third, the ABM does not yet include a complete strategic supply chain with distributors, queues, border diversion, and informal resale, although the scarcity state approximates availability. Fourth, health outcomes are reported as exposure because the Demographic and Health Survey is cross-sectional. Fifth, the simulation is partial equilibrium and does not model food-price inflation, wages, employment, exchange-rate feedback, or the alternative use of fiscal savings.

A stronger causal design would require staggered reform, territorial heterogeneity in implementation, repeated household surveys, or plausibly exogenous logistics shocks. With such variation, difference-in-differences and event studies could be combined with double/debiased machine learning for high-dimensional nuisance functions \parencite{Chernozhukov2018}. Generalized random forests could estimate heterogeneous effects under continuous or multivalued treatments \parencite{Athey2019}, and causal forests could estimate conditional compensation effects when a credible counterfactual exists \parencite{WagerAthey2018}. Until such data are available, machine learning should learn household states and heterogeneity for the ABM, not manufacture causality.

\begin{table}[H]
\centering
\caption{Assumption and identification registry}
\label{tab:registry}
{\footnotesize
\begin{tabularx}{\textwidth}{>{\raggedright\arraybackslash}p{2.45cm}>{\raggedright\arraybackslash}p{1.75cm}>{\raggedright\arraybackslash}p{2.55cm}>{\raggedright\arraybackslash}p{1.05cm}Y}
\toprule
Component & Status & Source & Risk & Mitigation \\
\midrule
LPG quantity & Estimated & Budget Survey, product 0452020101 & Medium & Department CV and survey reconciliation \\
Food vulnerability & Learned & Household Survey without target leakage & Medium & Ensemble, isotonic calibration, geographic holdout \\
Fuel affinity & Learned & Multicategory Household Survey models & Medium & Model comparison and disagreement \\
Price response & Calibrated & Scenario parameter & High & Adversarial grid and global sensitivity \\
Supply scarcity & \makecell[l]{Observed/\\forecast} & \makecell[l]{Monthly hydrocarbon\\series} & Medium & Temporal holdout and seasonal benchmark \\
Child health & Risk bridge & DHS 2023 & High & Report exposure, not causal clinical cases \\
Damage weights & Normative & Welfare function & High & Publish components and alternative weights \\
\bottomrule
\end{tabularx}
}
\end{table}

\section{Policy design, conclusions, and reproducibility}

The appropriate instrument depends on the binding constraint. If LPG is available and the problem is purchasing power, a targeted monthly energy transfer delivered through an existing channel minimizes administrative cost and allows consumption smoothing. If there is physical scarcity, queuing, hoarding, or a need to ration a subsidized quantity, a nominative digital voucher can be justified by traceability and quota enforcement despite its higher cost. If the household already cooks with firewood, dung, or other solid fuel, neither a fixed cash transfer nor an LPG voucher resolves the technological barrier: the policy requires a stove, cylinder, safety support, distribution access, and refill continuity. If a household includes a pregnant woman or young child, a mother-child component protects a high-return life-course window, but it does not replace an energy policy for poor households without children.

\begin{figure}[H]
\centering
\resizebox{0.94\textwidth}{!}{%
\begin{tikzpicture}[node distance=8mm and 12mm, every node/.style={font=\small}, q/.style={diamond,aspect=2.2,draw=navy,thick,fill=pale,align=center,text width=3.0cm}, a/.style={rounded corners,draw=teal,thick,fill=teal!8,align=center,text width=3.2cm,minimum height=11mm}, arr/.style={-{Latex[length=2.2mm]},thick}]
\node[q] (fuel) {Does the household use solid fuel?};
\node[a,below left=of fuel] (kit) {Clean-energy transition kit and support};
\node[q,below right=of fuel] (glp) {Does it use LPG and meet vulnerability criteria?};
\node[a,below left=of glp] (none) {No LPG component; assess other protection};
\node[q,below right=of glp] (short) {Is there physical scarcity or rationing?};
\node[a,below left=of short] (cash) {Monthly energy cash through an existing channel};
\node[a,below right=of short] (voucher) {Nominative, non-accumulating digital voucher};
\node[a,below=15mm of short] (mc) {Add mother-child protection when applicable};
\draw[arr] (fuel) -- node[left]{Yes} (kit); \draw[arr] (fuel) -- node[right]{No} (glp);
\draw[arr] (glp) -- node[left]{No} (none); \draw[arr] (glp) -- node[right]{Yes} (short);
\draw[arr] (short) -- node[left]{No} (cash); \draw[arr] (short) -- node[right]{Yes} (voucher);
\draw[arr] (kit) |- (mc); \draw[arr] (cash) |- (mc); \draw[arr] (voucher) |- (mc);
\end{tikzpicture}%
}
\caption{Institutional decision rule by type of constraint}
\label{fig:decision}
\source{Author's design.}
\end{figure}
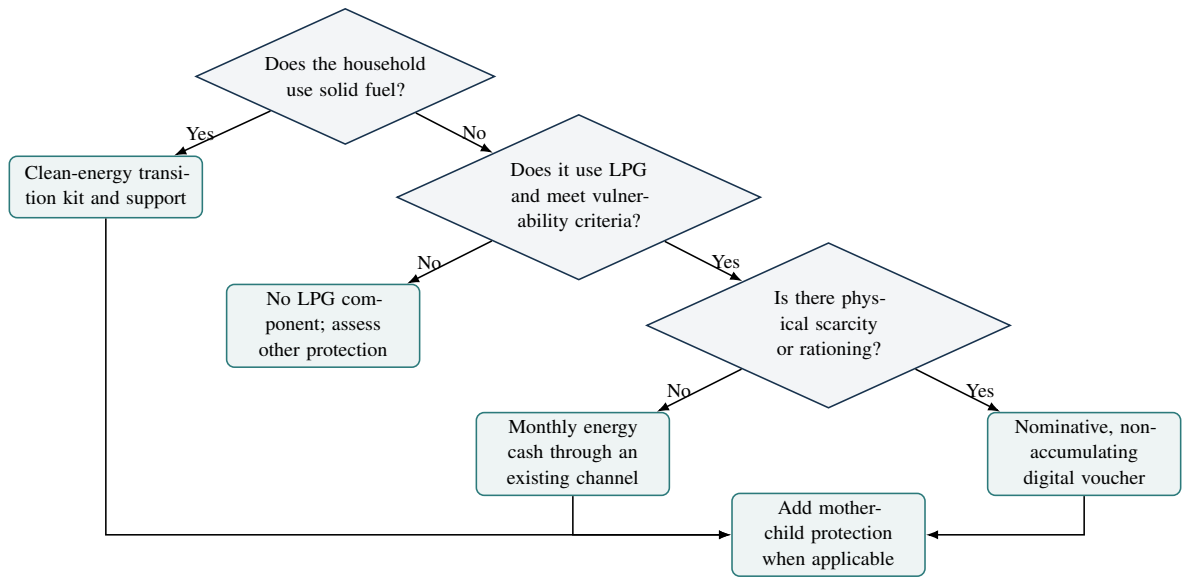

A prudent sequence has four elements. First, construct an interoperable registry using poverty, cooking fuel, and network-gas access, with an offline route for households without connectivity. Second, announce and pay compensation before the price adjustment to prevent a temporary liquidity loss, consistent with international experience in energy-subsidy reform \parencite{Mukherjee2023}. Third, adjust prices gradually while monitoring domestic sales, queues, refill access, and informal market premia. Fourth, publish eligibility rules, administrative costs, take-up, leakage, and inclusion-exclusion errors. Monthly cash aligns protection with the purchase flow; paying the entire annual amount in advance reduces transactions but increases the risk of early depletion and weakens the timing link with energy expenditure.

The policy emerging from the simulation is not immediate and universal subsidy removal. It is a conditional architecture: phase out the generalized price benefit, provide full-gap transitional compensation to vulnerable Q1-Q2 LPG households, reinforce mother-child protection, and finance clean-energy kits for households using solid fuels. Under the central scenario, this package preserves positive net fiscal savings while substantially reducing social damage relative to uncompensated removal. Cash dominates the voucher under normal physical availability; the voucher is reserved for scarcity and rationing. Universal LPG compensation is costly and weakly progressive because a large portion of the benefit goes to nonpoor households.

The analysis also has an epistemic conclusion. Microsimulation is strongest for accounting identities and distributional recalculation. Machine learning adds value by representing heterogeneity, but moderate out-of-sample performance requires restraint. The ABM organizes dynamics and counterfactual mechanisms, but every calibrated mechanism must remain visible and testable. Placebos and sensitivity analysis do not make the model causal; they reveal which internal claims survive falsification and which parameters dominate uncertainty. That distinction is essential for an auditable public-policy model.

The reproducibility package contains processed data, a variable dictionary, CSV and Markdown tables, PNG and PDF figures, serialized models, modular scripts, configuration templates, execution logs, methodological notes, and a SHA-256 manifest. A master execution script runs the complete pipeline, and a separate manuscript-graphics script regenerates the additional publication figures. Exact commands and software requirements are documented in the replication notes. Restricted raw microdata are not redistributed where licensing does not permit it; the lineage note records sources and transformations. The English deposit associated with this manuscript is \href{https://doi.org/10.5281/zenodo.21287432}{10.5281/zenodo.21287432}.

\begin{table}[H]
\centering
\caption{Final assessment of results and defensible uses}
\label{tab:assessment}
\begin{tabularx}{\textwidth}{>{\raggedright\arraybackslash}p{3.6cm}>{\raggedright\arraybackslash}p{2.8cm}Y}
\toprule
Result & Assessment & Defensible use \\
\midrule
Savings from the price differential & Strong, conditional & Fiscal planning under explicit price scenarios \\
Poverty reclassification & Strong as partial equilibrium & Comparison of compensation rules and thresholds \\
Cash-versus-voucher ranking & Robust under central costs & Institutional design when LPG is physically available \\
Individual demand magnitude & Moderate & National aggregation and sensitivity, not household auditing \\
Food-insecurity probabilities & Moderate & Vulnerability stratification \\
Solid-fuel transition pressure & Conditional & Stress testing and preventive prioritization \\
Child exposure & Exploratory & Risk signal, not a causal health effect \\
Optimal assignment & Conditional & Comparison of rules and budgets, not automated entitlement denial \\
\bottomrule
\end{tabularx}
\end{table}

\clearpage
\printbibliography[title={References}]

@online{INE2024EH,
  author       = {{National Institute of Statistics of Bolivia}},
  title        = {2024 Household Survey (EH 2024)},
  year         = {2025},
  organization = {National Institute of Statistics of Bolivia},
  url          = {https://anda.ine.gob.bo/index.php/catalog/163},
  urldate      = {2026-07-09}
}

@report{INE2019EPF,
  author      = {{National Institute of Statistics of Bolivia}},
  title       = {2015--2016 Family Budget Survey: Methodology and Results},
  institution = {National Institute of Statistics of Bolivia},
  location    = {La Paz},
  year        = {2019},
  url         = {https://www.ine.gob.bo/index.php/publicaciones/encuesta-de-presupuestos-familiares-2015-2016-metodologia-y-resultados/},
  urldate     = {2026-07-09}
}

@online{INE2025EDSA,
  author       = {{National Institute of Statistics of Bolivia}},
  title        = {2023 Demographic and Health Survey},
  year         = {2025},
  organization = {National Institute of Statistics of Bolivia},
  url          = {https://anda.ine.gob.bo/index.php/catalog/119},
  urldate      = {2026-07-09}
}

@online{INE2026Hydro,
  author       = {{National Institute of Statistics of Bolivia}},
  title        = {Hydrocarbons: Monthly Statistical Tables},
  year         = {2026},
  organization = {National Institute of Statistics of Bolivia},
  url          = {https://www.ine.gob.bo/index.php/estadisticas-economicas/hidrocarburos-mineria/hidrocarburo-cuadros-estadisticos/},
  urldate      = {2026-07-09}
}

@online{YPFB2026,
  author       = {{Yacimientos Petrolíferos Fiscales Bolivianos}},
  title        = {Products and Prices},
  year         = {2026},
  organization = {YPFB},
  url          = {https://www.ypfb.gob.bo/Nuestros_productos},
  urldate      = {2026-07-09}
}

@online{WorldBankWDI2026,
  author       = {{World Bank}},
  title        = {World Development Indicators},
  year         = {2026},
  organization = {World Bank},
  url          = {https://wdi.worldbank.org/},
  urldate      = {2026-07-09}
}

@article{Grimm2020,
  author  = {Grimm, Volker and Railsback, Steven F. and Vincenot, Christian E. and Berger, Uta and Gallagher, Cara and DeAngelis, Donald L. and Edmonds, Bruce and Ge, Jiaqi and Giske, Jarl and Groeneveld, Jürgen and Johnston, Alice S. A. and Milles, Alexander and Nabe-Nielsen, Jacob and Polhill, J. Gary and Radchuk, Viktoriia and Rohwäder, Maja-Sina and Stillman, Richard A. and Thiele, Jan C. and Ayllón, Daniel},
  title   = {The ODD Protocol for Describing Agent-Based and Other Simulation Models: A Second Update to Improve Clarity, Replication, and Structural Realism},
  journal = {Journal of Artificial Societies and Social Simulation},
  year    = {2020},
  volume  = {23},
  number  = {2},
  pages   = {7},
  doi     = {10.18564/jasss.4259}
}

@article{Muller2013,
  author  = {Müller, Birgit and Bohn, Friedrich and Dreßler, Gunnar and Groeneveld, Jürgen and Klassert, Christian and Martin, Romina and Schlüter, Maja and Schulze, Jule and Weise, Hanna and Schwarz, Nina},
  title   = {Describing Human Decisions in Agent-Based Models: ODD+D, an Extension of the ODD Protocol},
  journal = {Environmental Modelling \& Software},
  year    = {2013},
  volume  = {48},
  pages   = {37--48},
  doi     = {10.1016/j.envsoft.2013.06.003}
}

@article{Collins2024,
  author  = {Collins, Andrew J. and Koehler, Matthew and Lynch, Christopher J.},
  title   = {Methods That Support the Validation of Agent-Based Models: An Overview and Discussion},
  journal = {Journal of Artificial Societies and Social Simulation},
  year    = {2024},
  volume  = {27},
  number  = {1},
  pages   = {11},
  doi     = {10.18564/jasss.5258}
}

@article{Breiman2001,
  author  = {Breiman, Leo},
  title   = {Random Forests},
  journal = {Machine Learning},
  year    = {2001},
  volume  = {45},
  number  = {1},
  pages   = {5--32},
  doi     = {10.1023/A:1010933404324}
}

@article{Friedman2001,
  author  = {Friedman, Jerome H.},
  title   = {Greedy Function Approximation: A Gradient Boosting Machine},
  journal = {The Annals of Statistics},
  year    = {2001},
  volume  = {29},
  number  = {5},
  pages   = {1189--1232},
  doi     = {10.1214/aos/1013203451}
}

@inproceedings{Niculescu2005,
  author    = {Niculescu-Mizil, Alexandru and Caruana, Rich},
  title     = {Predicting Good Probabilities with Supervised Learning},
  booktitle = {Proceedings of the 22nd International Conference on Machine Learning},
  year      = {2005},
  pages     = {625--632},
  publisher = {ACM},
  doi       = {10.1145/1102351.1102430}
}

@article{Chernozhukov2018,
  author  = {Chernozhukov, Victor and Chetverikov, Denis and Demirer, Mert and Duflo, Esther and Hansen, Christian and Newey, Whitney and Robins, James},
  title   = {Double/Debiased Machine Learning for Treatment and Structural Parameters},
  journal = {The Econometrics Journal},
  year    = {2018},
  volume  = {21},
  number  = {1},
  pages   = {C1--C68},
  doi     = {10.1111/ectj.12097}
}

@article{Athey2019,
  author  = {Athey, Susan and Tibshirani, Julie and Wager, Stefan},
  title   = {Generalized Random Forests},
  journal = {The Annals of Statistics},
  year    = {2019},
  volume  = {47},
  number  = {2},
  pages   = {1148--1178},
  doi     = {10.1214/18-AOS1709}
}

@article{WagerAthey2018,
  author  = {Wager, Stefan and Athey, Susan},
  title   = {Estimation and Inference of Heterogeneous Treatment Effects Using Random Forests},
  journal = {Journal of the American Statistical Association},
  year    = {2018},
  volume  = {113},
  number  = {523},
  pages   = {1228--1242},
  doi     = {10.1080/01621459.2017.1319839}
}

@article{Saltelli2010,
  author  = {Saltelli, Andrea and Annoni, Paola and Azzini, Ivano and Campolongo, Francesca and Ratto, Marco and Tarantola, Stefano},
  title   = {Variance Based Sensitivity Analysis of Model Output: Design and Estimator for the Total Sensitivity Index},
  journal = {Computer Physics Communications},
  year    = {2010},
  volume  = {181},
  number  = {2},
  pages   = {259--270},
  doi     = {10.1016/j.cpc.2009.09.018}
}

@article{Masera2000,
  author  = {Masera, Omar R. and Saatkamp, Barbara D. and Kammen, Daniel M.},
  title   = {From Linear Fuel Switching to Multiple Cooking Strategies: A Critique and Alternative to the Energy Ladder Model},
  journal = {World Development},
  year    = {2000},
  volume  = {28},
  number  = {12},
  pages   = {2083--2103},
  doi     = {10.1016/S0305-750X(00)00076-0}
}

@report{WHO2014,
  author      = {{World Health Organization}},
  title       = {WHO Guidelines for Indoor Air Quality: Household Fuel Combustion},
  institution = {World Health Organization},
  location    = {Geneva},
  year        = {2014},
  isbn        = {9789241548885},
  url         = {https://www.who.int/publications/i/item/9789241548885},
  urldate     = {2026-07-09}
}

@article{Hidrobo2014,
  author  = {Hidrobo, Melissa and Hoddinott, John and Peterman, Amber and Margolies, Amy and Moreira, Vanessa},
  title   = {Cash, Food, or Vouchers? Evidence from a Randomized Experiment in Northern Ecuador},
  journal = {Journal of Development Economics},
  year    = {2014},
  volume  = {107},
  pages   = {144--156},
  doi     = {10.1016/j.jdeveco.2013.11.009}
}

@article{Cunha2014,
  author  = {Cunha, Jesse M.},
  title   = {Testing Paternalism: Cash versus In-Kind Transfers},
  journal = {American Economic Journal: Applied Economics},
  year    = {2014},
  volume  = {6},
  number  = {2},
  pages   = {195--230},
  doi     = {10.1257/app.6.2.195}
}

@report{Mukherjee2023,
  author      = {Mukherjee, Anit and Okamura, Yuko and Gentilini, Ugo and Gencer, Defne and Almenfi, Mohamed and Kryeziu, Adea and Montenegro, Miriam and Umapathi, Nithin},
  title       = {Cash Transfers in the Context of Energy Subsidy Reforms: Insights from Recent Experience},
  institution = {World Bank},
  series      = {Energy Subsidy Reform in Action},
  location    = {Washington, DC},
  year        = {2023},
  url         = {https://documents1.worldbank.org/curated/en/099062923170018606/pdf/P17658505cf5310870baf305828791be2a8.pdf},
  urldate     = {2026-07-09}
}

\end{document}